# Pulse Profiles of Swift J1626.6+5156

Baykal, A.*, İnam, S.Ç.† and İçdem, B.*

*METU, Physics Department, 06531 Ankara, Turkey
†Başkent University, Department of Electrical and Electronics Engineering, 06530 Ankara, Turkey

**Abstract.** In this paper, we analyzed pulse profiles of Swift J1626.6+5156 using the lightcurves from RXTE-PCA observations between MJD 53724 (just after the outburst) and MJD 55113 and a Chandra-ACIS dataset on MJD 54897 with a 20 ks exposure. We found that pulse profiles show morphological variations and pulsations do not cease even $\sim$ 1200 days after the outburst. Despite these variations, we did not find any significant variation in the pulsed fraction with decreasing X-ray flux.

**PACS:** 98

## INTRODUCTION

Swift J1626.6+5156 is a transient accretion powered pulsar with a spin period of $\sim$15 s (Palmer et al. 2005; Markwardt et al. 2005) which was first detected on December 2005 with the Swift Burst Alert Telescope (BAT) (Palmer et al 2005). Following the 2005 outburst of the source, Reig et al. (2008) found a $\sim$ 450 s X-ray flare, during which the pulsed fraction increased up to $\sim$ 70%. After the flare, the average count rate and the pulsed fraction were restored to their original values.

Proposed optical companion of the source was found to show strong H$\alpha$ emission, which is an indication of a Be star (Negueruela & Marco, 2006). As the infrared magnitudes of the companion is rather large for a Be star (J=13.5, H=13, K=12.6; Rea et al. 2006), the system is considered to be an unusual Be/X$-$ray binary system.

Baykal et al. (2010) studied long term monitoring RXTE (Rossi X-ray Timing Explorer) - PCA (Proportional Counter Array) observations of the source between MJD 53724 and 54410. They reported the orbital period to be 132.89 days. Baykal et al. (2010) also constructed long term frequency history of Swift J1626.6+5156 and found that the timescale of the X-ray modulations varied. İçdem et al. (2011) extended timing analysis of Baykal et al. (2010) with the new analysis of RXTE-PCA data until MJD 55113 together with the Chandra-ACIS dataset with a 20 ks exposure. İçdem et al. (2011) also studied X-ray spectral variations of the source and found that spin-up rate of the source is correlated with the X-ray flux.

In this paper, we study pulse profile evolution of the source with RXTE-PCA observations between MJD 53724 and MJD 55113 together with a Chandra-ACIS dataset on MJD 54897 with a 20 ks exposure.

## OBSERVATIONS

We analyzed data from Proportional Counter Array (PCA) onboard RXTE (Jahoda et al 1996) of Swift J1626.6+5156 between MJD 53724 and MJD 55113 with a total exposure of $\sim 449$ ks, divided into 411 observations with exposures between $\sim 1$ ks and $\sim 2$ ks. This work uses results of the timing and spectral analysis of RXTE-PCA data between MJD 53724 and 54410 (Baykal et al. 2010; İçdem et al. 2011) and between MJD 54410 and MJD 55113 (İçdem et al. 2011).

In addition to the RXTE-PCA data, we also used a Chandra AXAF CCD Imaging Spectrometer (ACIS; Garmire et al. 2003) observation of Swift J1626.6+5156 on MJD 54897 with an exposure of 20ks. This observation contains ACIS-S FAINT TE(timed exposure) mode data.

## ANALYSIS AND RESULTS

To obtain pulse profiles from RXTE-PCA observations, we used background subtracted and solar system barycenter corrected 3-20 keV RXTE-PCA lightcurves obtained by İçdem et al. (2011) with a time resolution of 0.375s. Pulse profiles were obtained by folding the lightcurves with the corresponding pulse frequencies (see İçdem et al. 2011). Sample RXTE-PCA pulse profiles were presented in Figure 1.

We calculated pulsed fraction of these pulse profiles using the definition $(F_{max} - F_{min})/(F_{max} + F_{min})$ where $F_{max}$ and $F_{min}$ are the highest and lowest fluxes of the phase bins. In order to obtain flux dependence of the pulsed fraction, we used unabsorbed 3-20 keV X-ray flux values from the X-ray spectral analysis results of İçdem et al. (2011). In Figure 2, we present pulsed fraction values as a function of X-ray flux.

Using Chandra-ACIS observations, we obtain an overall pulse profile from the 20ks long 0.3-8 keV lightcurve of the source with a time resolution of 0.44s by folding this lightcurve with the pulse frequency of $(6.52059 \pm 0.00075) \times 10^{-2}$ Hz (see Figure 1). This pulse frequency value was obtained by İçdem et al. (2011) using cross-correlation of individual pulses obtained from the lightcurve.

## DISCUSSION

From Figure 1, it is shown that pulses from Swift J1626.6+5156 do not cease completely in case X-ray flux of the source decreases significantly. Pulse profile obtained from the Chandra observation especially indicates that pulses from the source do not cease even after $\sim 1200$ days from the outburst. However, pulse shapes vary considerably and for the observations when the source is not bright, we occasionally see double peaks in the pulse profile. This should be an indication of an X-ray flux related accretion geometry change.

It is also important to note that pulsed fraction of the source does not show any clear correlation with the unabsorbed X-ray flux (see Figure 2). For accretion powered pulsars, pulsed fraction changes can be interpreted as a sign of accretion geometry changes. Lack of this correlation despite the presence of pulse profile variations distinguishes Swift

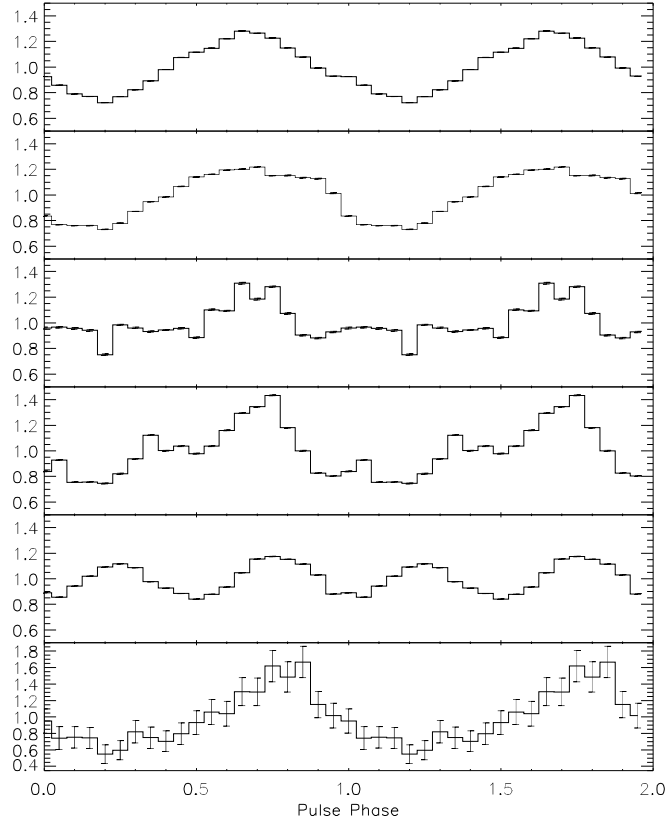

**FIGURE 1.** Sample pulse profiles obtained from RXTE PCA and Chandra ACIS observations. The first five of these pulse profiles were obtained from ∼ 2 ks long 3-20 keV RXTE-PCA lightcurves at (from top to bottom) MJD ∼ 53756.6, MJD ∼ 53935.8, MJD ∼ 54126.9, MJD ∼ 54350.3 and MJD ∼ 54601.7. The lowermost pulse profile was obtained by İçdem et al. (2011) from 0.3-8 keV Chandra-ACIS lightcurve obtained from the observation on MJD 54897.

J1626.6+5156 from other transient accretion powered pulsars 2S 1417-62 (İnam et al. 2004) and XMMU J054134.7-682550 (İnam et al. 2009).

## ACKNOWLEDGMENTS

We acknowledge research project TBAG 109T748 of the Scientific and Technological Research Council of Turkey (TÜBİTAK).

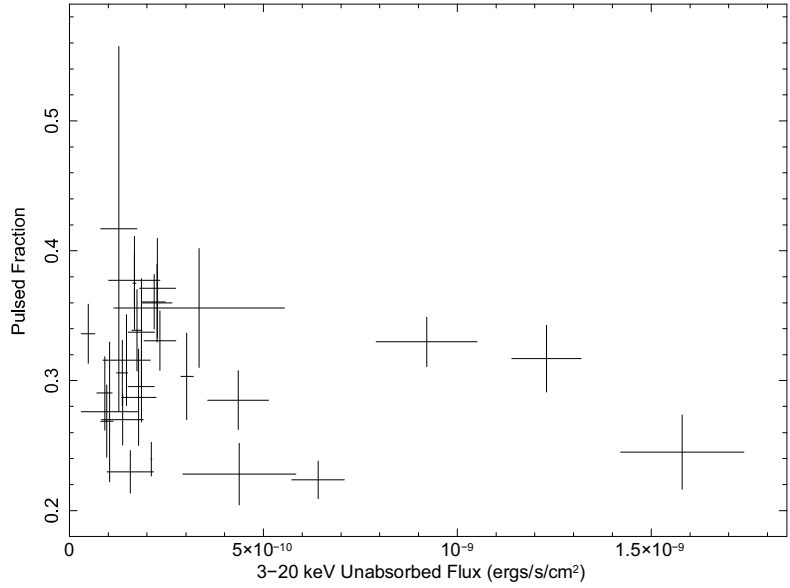

**FIGURE 2.** Pulsed fraction as a function of 3-20 keV unabsorbed X-ray Flux. X-ray flux values obtained from X-ray spectral analysis of the source by İçdem et al. (2011).